\documentclass[twocolumn,showpacs]{revtex4-1}
\usepackage{subcaption} \usepackage{graphicx} \usepackage{amsmath}
\graphicspath{{./fig/}}
\usepackage{url}
\usepackage{color}
\newcommand{\red}[1]{\textcolor{red}{{#1}}}

\begin{document} 

\title{\red{View \& Perspective, Frontiers of Physics}\\ \vspace{1cm}
Tasting Nuclear Pasta Made with Classical Molecular Dynamics Simulations}

\author{Bao-An Li}

\affiliation{Department of Physics and Astronomy, Texas A\&M University-Commerce, TX 75429-3011, USA}


\begin{abstract}
Nuclear clusters or voids in the inner crust of neutron stars were predicted to have various shapes collectively nicknamed nuclear pasta. The recent review in Ref. \cite{Lopez1} by L\'opez, Dorso and Frank summarized their systematic investigations into properties especially the morphological and thermodynamical phase transitions of the nuclear pasta within a Classical Molecular Dynamics model, 
providing further stimuli to find more observational evidences of the predicted nuclear pasta in neutron stars. 
\end{abstract}

\maketitle
Known as the densest visible object in the Universe, neutron stars have many mysterious properties to be understood. From the external magnetic field through the surface and crust to the core of neutron stars,
there are many fundamental questions to be addressed \cite{NAP2011,NAP2012}. Thanks to the recent advances especially in radio, X-ray and gravitational wave observations of both isolated neutron stars and their mergers, much 
progresses have been made in recent years in understanding the maximum mass, radius and tidal deformation of neutron stars. These global observables provided some of the much needed constraints on various 
theories about the equation of state, internal structure and composition of neutron stars.  However, because of the scarcity of data available especially those from the un-distorted messengers directly from the interior of neutron stars and the model dependence of their interpretations, many mysteries associated with neutron stars remain to be resolved. In fact, to understand the nature of neutron stars and dense neutron-rich matter has been a long-standing and shared goal of both astrophysics and nuclear physics communities \cite{LRP2015,NuPECC}. 

Among the main challenges, one important task is to understand the structure, content and formation mechanism of neutron stars. People have imagined that neutron stars have several layers with distinct features from the atmosphere, envelope, crust to the outer and inner core in an analogy to the structure of earth. The outer core of neutron stars is speculated to be some kind of fluid consisting of neutron, protons, electrons, muons, hyperons, etc. As the density of this fluid decreases towards the crust, the system is expected to enter the so-called spinodal region where the matter (either pure nucleonic matter of neutrons and protons \cite{Siemens83} or neutron star matter \cite{Lat00}) is dynamically unstable against the growth of small density fluctuations when the temperature is sufficiently low. Consequently, bubbles/voids or droplets/clusters are expected to develop depending on the trajectory of the system before entering the spinodal region, marking a transition between the uniform core and the inhomogeneous crust. This transition is expected to happen in the inner crust and the exact transition density has been found to depend sensitively on the characteristics of nuclear symmetry energy encoding the energy necessary to make nuclear matter more neutron rich, see, e.g., Refs. \cite{Li19,JXu,Newton12}. Very interestingly,  because of the competition between the surface tension and long-range Coulomb force, the bubbles and nuclear clusters may have various shapes: from the spherical void near the core, through the cylindrical void to the plates, cylinders and spheres of nucleons towards the surface of neutron stars. Because of their similarity to meatballs, spaghetti, lasagna, macaroni, and Swiss cheese, respectively, these structures were collectively nicknamed nuclear pasta. For an earlier review of
the physics of nuclear pasta, see, Ref.\cite{Pethick}. Since the pioneering works of Ravenhall \emph{et al.}~\cite{Rav} and Hashimoto \emph{et al.}~\cite{Has}, besides the above five classical shapes, several other shapes, such as the gyroid, double-diamond \cite{Ken}, sponge-like \cite{Dorso} and parking-garage like \cite{Hor1} structures have been predicted using various approaches, see, e.g., Refs. \cite{Wil85,Oya93,Lor93,Che97,Wat00,Wat02,Wat03,Mar98,Kid00,Hor04}. Interestingly, similar shapes have been found in nano materials and/or biological systems. 

The various shapes of the nuclear voids and clusters correspond to the local minima of potential energies separated by energy barriers. Often, the energy minima are very close to each other. The realization of various shapes are thus somewhat model dependent and the transition from one shape to another depends sensitively on the nuclear physics inputs, such as the nuclear equation of state especially its symmetry energy term \cite{Newton12,Newton09,Bao,Kaz20,Xia}.  Both static and dynamic approaches using various nuclear interactions and assumptions about the composition of neutron stars have been used in studying the formation, properties and phase transition of nuclear pasta by many groups over the last three decades. The recent review in Ref. \cite{Lopez1} by L\'opez, Dorso and Frank summarized their systematic investigations into properties especially the morphological and thermodynamical phase transitions of the nuclear pasta within a Classical Molecular Dynamics model. Using several morphologic and thermodynamic 
tools, they characterized the morphology of the emerging structures in neutron star crust by varying the temperature, density, neutron to proton ratio with and without considering electrons. They constructed the phase diagrams for both nucleonic matter and neutron star matter. They also investigated the isospin (neutron to proton ratio) dependence of the critical points and morphologic phase transition as well as the symmetry energy of clustered matter. Bearing in mind the possible model dependence, effects of some poorly known but necessary nuclear physics inputs, finite size effects in simulating infinite matter using finite-sized unit cells with periodic boundary condition and the lack of quantum effects, their results obtained with the Classical Molecular Dynamics simulations are certainly stimulating and useful for further exploring the interesting physics of nuclear pasta. 

As the nuclear pasta may actually exist inside the core of supernovae and the crust of neutron stars, the physics of nuclear pasta is important for understanding some astrophysical observations \cite{Cham,Newton14,Chuck}. For example, supernova explosions, protoneutron star cooling mechanism and the associated neutrino transport \cite{Hor04,Chuck,Rog,All,Newton13,Sc}, pulsar glitches \cite{Wat17,Josh}, quadrupole deformation or formation of mountains on neutron stars and the associated gravitational wave emission \cite{Cap,Pet20,Bis,And}, frequencies of torsional oscillations of neutron stars and the associated mechanism for generating quasi-periodic oscillations in the tails of light curves of giant flares from soft gamma-ray repeaters \cite{Gea,Sot}, all depend strongly on properties of neutron star crust especially whether the nuclear pasta is considered or not. Moreover, the crustal properties are also important for determining the r-mode stability window of super-fast pulsars \cite{Wen,Isaac}. They are also critical for determine whether GW190814's second component of mass (2.5-2.67)M$_{\odot}$ discovered by LIGO/VIRGO \cite{GW} very recently is the lightest black hole or the most massive and fastest rotating neutron star observed so far \cite {Most,Zhang} as pointed out in Ref. \cite{rmode}.  Of course, the main challenge is to find unique signatures of the nuclear pasta as most of the astrophysical phenomena mentioned above have alternative explanations without considering the nuclear pasta in the inner crust of neutron stars.  Hopefully, the review in Ref. \cite{Lopez1} by L\'opez, Dorso and Frank will stimulate more work in this direction.

BALI is supported in part by the U.S. Department of Energy, Office of Science, under Award Number DE-SC0013702 and the CUSTIPEN (China-U.S. Theory Institute for Physics with Exotic Nuclei) under the US Department of Energy Grant No. DE-SC0009971.

\end{document}